\documentclass[conference]{IEEEtran}
\IEEEoverridecommandlockouts
\usepackage{cite}
\usepackage{amsmath,amssymb,amsfonts}
\usepackage{algorithmic}
\usepackage{graphicx}
\usepackage{textcomp}
\usepackage{xcolor}

\usepackage{nameref}

\usepackage[utf8]{inputenc}
\usepackage{float}

\def\BibTeX{{\rm B\kern-.05em{\sc i\kern-.025em b}\kern-.08em
    T\kern-.1667em\lower.7ex\hbox{E}\kern-.125emX}}

\makeatletter
\def\ps@IEEEtitlepagestyle{%
  \def\@oddfoot{\mycopyrightnotice}%
  \def\@evenfoot{}%
}
\def\mycopyrightnotice{%
  {\footnotesize The copyright belongs to me!\hfill}% <--- Change here
  \gdef\mycopyrightnotice{}% just in case
}
\def\ps@IEEEtitlepagestyle{%
  \def\@oddfoot{\mycopyrightnotice}%
  \def\@evenfoot{}%
}
\def\mycopyrightnotice{%
  {\footnotesize 978-1-5386-8375-0/18/\$31.00 \copyright 2018 IEEE \hfill}% <--- Change here
  \gdef\mycopyrightnotice{}% just in case
}
    
\begin{document}

%\IEEEoverridecommandlockouts
%    \IEEEpubid{\makebox[\columnwidth]{978-1-5386-8375-0/18/\$31.00 \copyright 2018 IEEE \hfill} \hspace{\columnsep}\makebox[\columnwidth]{ }}
%

\title{Text Mining over Curriculum Vitae of Peruvian Professionals using Official Scientific Site DINA}

\author{
\IEEEauthorblockN{Josimar Edinson Chire Saire}
\IEEEauthorblockA{\textit{Institute of Mathematics and Computer Science (ICMC)} \\
\textit{University of São Paulo (USP)}\\
São Carlos, SP, Brazil \\
jecs89@usp.br}
\and
\IEEEauthorblockN{Honorio Apaza Alanoca}
\IEEEauthorblockA{\textit{National University of Moquegua, Ilo}\\
\textit{National University of Moquegua,}\\
Ilo, Moquegua, Peru \\
honorio.apz@gmail.com}
}

\maketitle

\begin{abstract}

During the last decade, Peruvian government started to invest and promote Science and Technology through Concytec(National Council of Science and Technology). Many programs are oriented to support research projects, expenses for paper presentation, organization of conferences/ events and more. Concytec created a National Directory of Researchers(DINA) where professionals can create and add curriculum vitae, Concytec can provide official title of Researcher following some criterion for the evaluation. The actual paper aims to conduct an exploratory analysis over the curriculum vitae of Peruvian Professionals using Data Mining Approach to understand Peruvian context.

\end{abstract}


\section{Introduction}
In the last decade, information and communication technology has innovated considerably, in the field of administration and selection of human resources in companies it has also evolved, Job applicants send their Curriculum Vitae (CV) through the Web or send them directly to a company.As an application, text mining is feasible for commercial use for the creation of knowledge, from the point of view of transformation of unstructured data \cite{10.1145/312129.312299}, which assisted with a rational criterion of the human being. The e-procurement area is facing a growing number of these documents that are in different formats and contain a large amount of information\cite{5587018}. The process of transforming unstructured candidate data into knowledge graphs has become a major challenge in machine learning \cite{10.1145/3159652.3162011}. A common problem in the academic field is to select professionals with a good research and development R\&D profile.

Most of the time, identifying potential job candidates is an expensive and time-consuming task for human resources divisions \cite{sandanayake}.In the academic field there is a need to select the best candidate, this problem is very common in Peruvian universities.In the research \cite{Santana2017DataSA} an exploratory study of the information from Lattes, a specialized social network of researchers from Brazil, was carried out and presents a new approach to the analysis of regulated groups. The visualization components allow geographic exploration of collections, interpretation of the evolution of the topic \cite{10.1145/3127526.3127529}, At present it is a very important component to visualize, analyze and interpret the behavior of data that can express the inclination of research lines, languages, professions, among others.
Recently, Artificial Intelligence has been successfully exploited and tools based on Data Mining, Multi-Agent Systems and Knowledge Representation Approaches (Ontologies)\cite{10.1145/3301551.3301579}.\\
In context Peruvian in 2018, the investment reached 160 million soles and the amounts have advanced year after year  \cite{andina}.The figure responds to the investment of Concytec and the National Fund for Scientific, Technological and Technological Innovation Development.\\
In Peru, an investigation of recommendation of resumes based on relevance of terms was carried out, natural language processing techniques and text mining were applied \cite{alanoca2020curriculum}. However, there is no study carried out exclusively in the academic field, while there is information about the researchers registered in the National Directory of Researchers (DINA).
Therefore, the present investigation focuses on making an exploratory investigation and analysis of the data contained in curriculms vitae of the National Director of Researchers. In addition, it seeks that the results obtained can help to have a clearer picture of who evaluated the research and technological development in Peru.

\section{Proposal}

The present paper is an exploratory study of Peruvian Professionals using available Curriculum Vitae, following the next steps:

\begin{itemize}
    \item Selecting the scope
    \item Find the relevant terms to search
    \item Preprocessing
    \item Visualization
\end{itemize}

\subsection{Selecting the scope}

At the beginning, the motivation was related to study and understand Peruvian context about Research topic. Then, considering the existence of website gathering professional information in many countries, i.e. cv lattes(Brazil). The source of data is official website of Concytec.

\subsection{Find the relevant terms to search}

Considering the structure of website for each professional, html structure is analyzed to extract relevant data for the posterior phases. This step can uses $div\_class$ or $html\_tag$. 

\subsection{Preprocessing Data}

The collected data has text format, then to make it readable for next step, the next steps are considered:

\begin{itemize}
    \item Convert text to lowercase
    \item Remove non alphanumeric characters
    \item Remove stopwords
    \item Remove custom stopwords, i.e. experiencia(experience), inicio(start)
\end{itemize}

\subsection{Visualization}

The study is focusing on the analysis of Academic Information, Profesional Experience, Scientific Publications and Languages. Then, many filtering/selections steps related are performed, i.e. bachiller(bachelor), maestria(masters) and so on. Besides, cloud of words are presenting to visualize the frequency of terms.

\section{Results}

The next graphics are the result of the experiments on dataset. Subsection \ref{aca} is presenting Academic Information, subsection \ref{pro} presenting Professional Experience and Scientific Publications. Subsection \ref{lan} present information about languages, and inside of this part, \ref{subsubsec:per} presents information about Peruvian Languages.

\subsection{Description of dataset}

\begin{itemize}
    \item A collection of 25,000 registers were selected for this analysis.
    \item From this registers, only 14,504 has valid information.
    \item Fields: academic information(alphanumeric), professional experience(alphanumeric), scientific publications(alphanumeric), languages(alphanumeric)
    \item There are null values in some places, two options are possible: professional does not add any information or does not have any to add in one specific field.
\end{itemize}

\subsection{Academical Information }
\label{aca}

\begin{figure*}[hbpt]
\centerline{
\includegraphics [width=0.43\textwidth]{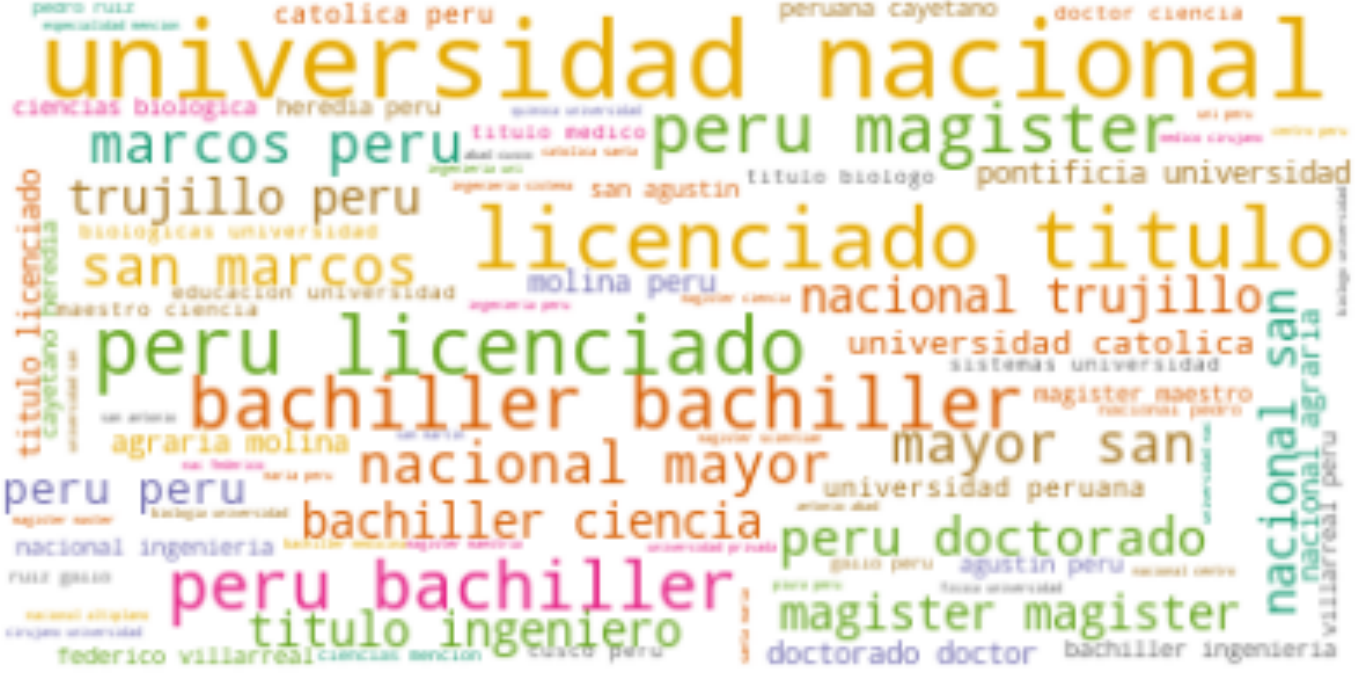}
\includegraphics [width=0.43\textwidth]{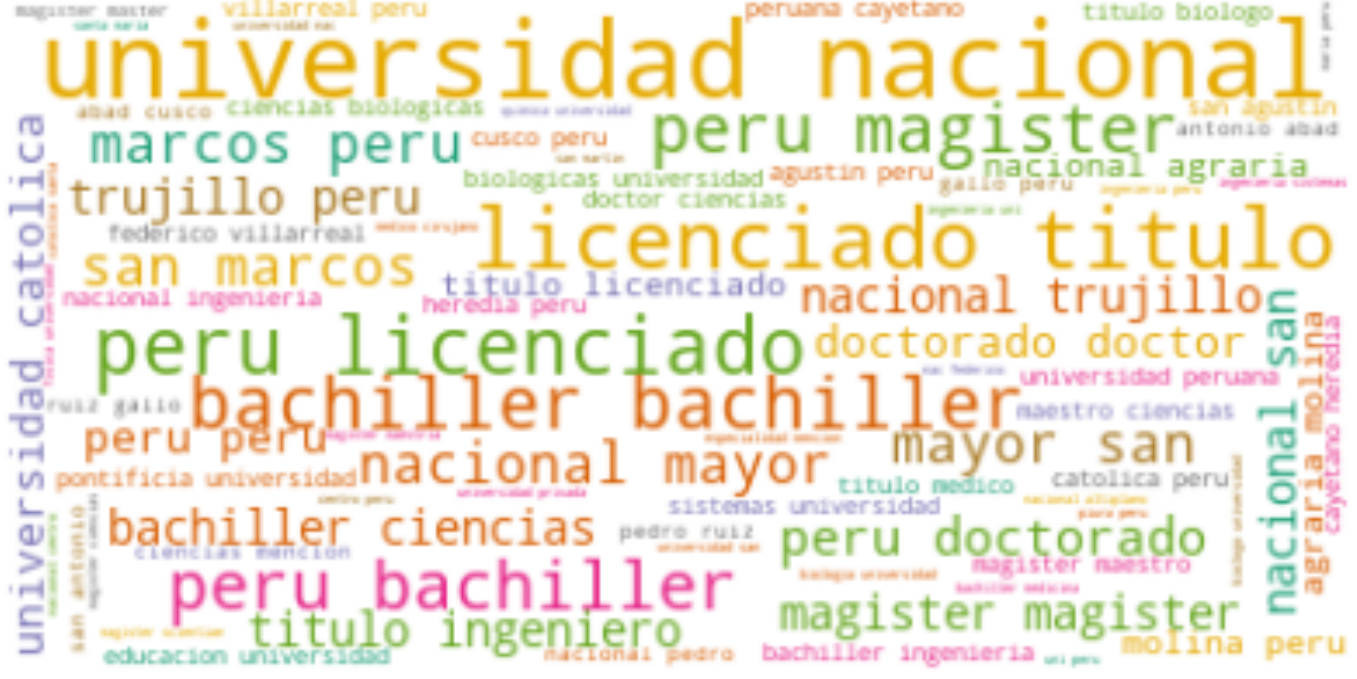}
}
\caption{Cloud of Words related to Academical Information (1-gram,2-gram)}
\label{fig:cw_acad}
\end{figure*}

Graphic \ref{fig:cw_acad} presents a cloud of words of the Academical Information, it is remarkable to see universidad nacional(national university), peru bachiller(peru bachelor). Then, first insight peruvian website has professionals who studied mainly in national universities. Peruvian national universities, are free of charge, this situation is different in other countries of South America, i.e. Chile. 

An exploratory query to data is performed to know if there is more countries where professionals studied bachelor, master or phd degrees. Table  \ref{tab:studies} presents the distribution in South America, where people got his degree respectively. A previour affirmation related to Peruvian universities is conffirmed, besides Brasil, Chile, Argentina are the top three in South American countries. At the other hand, considering the number of ocurrences, it is possible people got his degree outside of South America. Checking, column 'Bachelor', 'Master', 'Phd', the next proportions are calculated: 54\%, 77\%, 28\%. This last proportions indicates the potential of professional to work on Industry or Academy, most than 70\% has a Master Degree and almost one third holds a Phd degree.

\begin{table}[hbpt]
\caption{Distribution of Countries where professional got a degree}
\label{tab:studies}
\centering
\begin{tabular}{|c|c|c|c|c|}
\hline
\textbf{Country} & \textbf{Ocurrences} & \textbf{Bachelor} & \textbf{Master} & \textbf{Phd}  \\ \hline
Peru             & 11795               & 7147              & 9940            & 3416          \\ \hline
Argentina        & 101                 & 48                & 89              & 49            \\ \hline
Bolivia          & 12                  & 2                 & 11              & 3             \\ \hline
Brasil           & 681                 & 455               & 773             & 438           \\ \hline
Chile            & 272                 & 169               & 273             & 134           \\ \hline
Colombia         & 58                  & 17                & 59              & 22            \\ \hline
Ecuador          & 29                  & 8                 & 28              & 10            \\ \hline
Paraguay         & 7                   & 5                 & 5               & 2             \\ \hline
Uruguay          & 4                   & 0                 & 4               & 3             \\ \hline
Venezuela        & 23                  & 12                & 23              & 15            \\ \hline
\textbf{Total}   & \textbf{12982}      & \textbf{7863}     & \textbf{11205}  & \textbf{4092} \\ \hline
\end{tabular}
\end{table}

\subsection{Professional Experience, Scientific Production}
\label{pro}

\begin{figure*}[hbpt]
\centerline{
\includegraphics [width=0.43\textwidth]{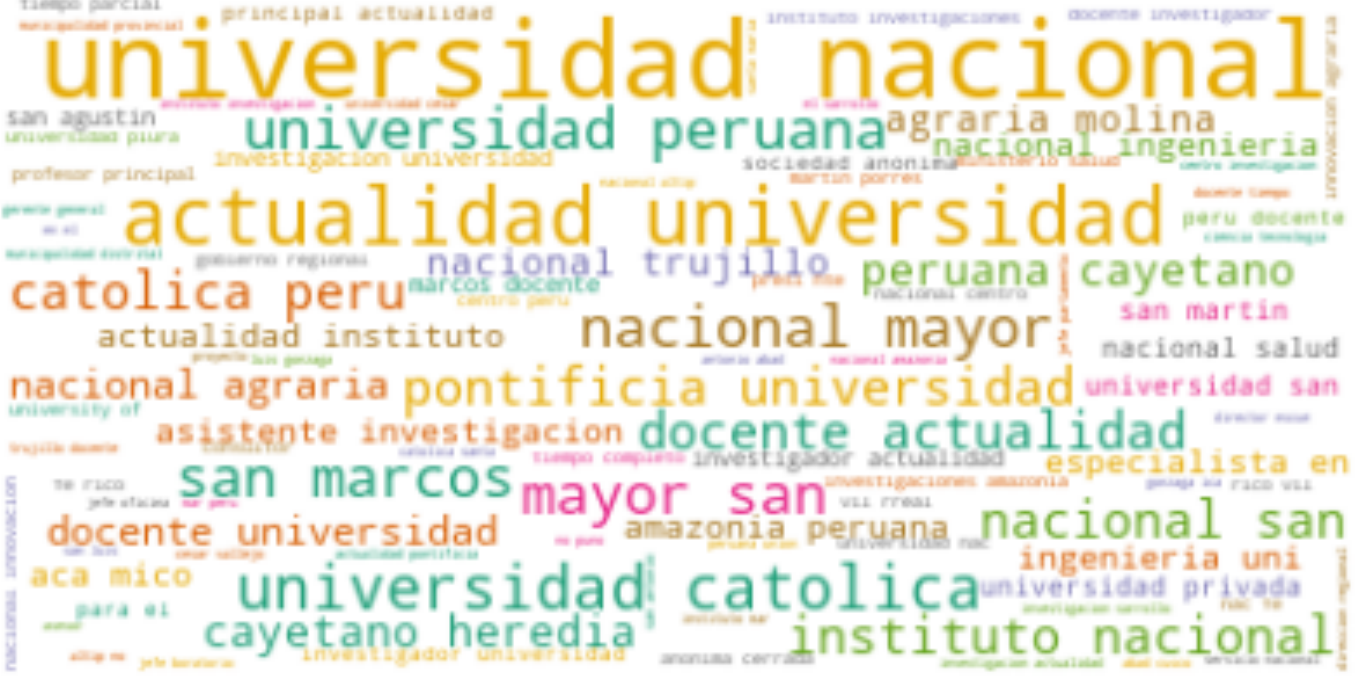}
\includegraphics [width=0.43\textwidth]{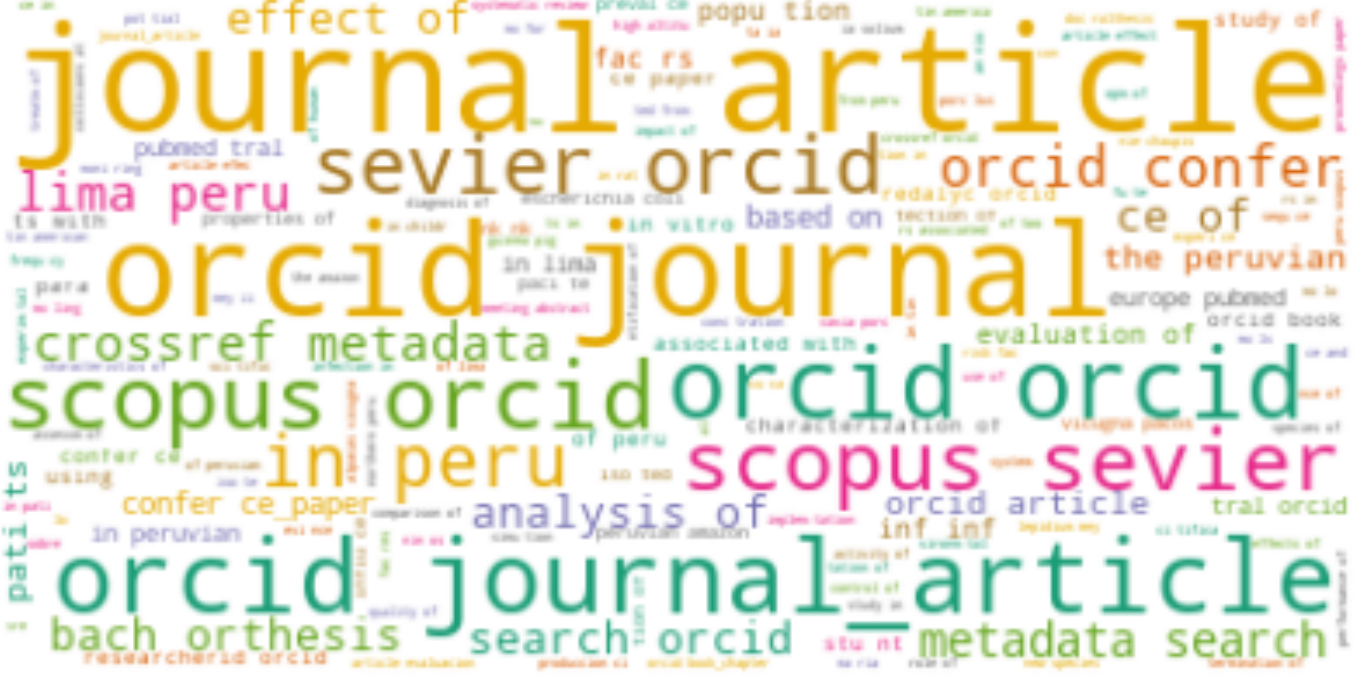}
}
\caption{Cloud of Words related to Profesional Experience and Scientific Production}
\label{fig:cw_profvssci}
\end{figure*}

This subsection presents graphic Fig. \ref{fig:cw_profvssci} to present where people worked or works now. Left side, shows terms: 'universidad'(university), 'nacional'(national), 'actualidad'(actualidad), 'docente(docente)', 'instituto'(institute), those words indicates professional are working in Academic Centers(university, institute), working as teachers and most of them in national universities.

Right side of the graphic, present cloud of words about Scientific Production, terms are: 'journal', 'article', 'orcid', 'scopus', 'sevier'. These words are related to scientic publications(papers) in conference/journals with indexation Scopus, El Sevier. Besides, Orcid is a unique code to identify researcher and their contributions in personal website. Then, it is possible to add or import all relevant information to DINA website.

\subsection{Languages}
\label{lan}

Considering results of subsection \ref{aca}, \ref{pro}, this subsection pretends to show how many languages or how is the level of languages of professional then this can be a way to export/show capacity through publications, talks, collaboration projects and more.

Most of the conferences/journals as English language as requirement, then it is necessary to write/read/speak it. Table \ref{tab:eng}, can express the level of reading/speaking/writing using scale of basic/intermediate/advanced/superior advanced. Science, Technology are evolving everyday, therefore is necessary to read how this process happens, reading skill is vital 9,321 (64.26\%). Write a publication/report or any scientific/industry document involves writing skill, 7,375 (50.84\%), finally to communicate in conferences presentations speaking is necessary, 8270 (57.02\%).

\begin{table}[hbpt]
\label{tab:eng}
\caption{English Level}
\centering
\begin{tabular}{|c|c|c|c|c|}
\hline
\textbf{English Level} & \textbf{Reading} & \textbf{Speaking} & \textbf{Writing} & \textbf{Total\_fil} \\ \hline
Basic                  & 2174             & 4120              & 3225             & \textbf{9519}       \\ \hline
Intermediate           & 4445             & 4194              & 4728             & \textbf{13367}      \\ \hline
Advanced               & 3722             & 2293              & 2664             & \textbf{8679}       \\ \hline
Superior Advanced      & 1154             & 888               & 878              & \textbf{2920}       \\ \hline
\textbf{Total\_col}    & \textbf{11495}   & \textbf{11495}    & \textbf{11495}   &                     \\ \hline
\end{tabular}
\end{table}

Brazil is one of the biggest countries in South America, official language is Portuguese. Then, half of South America speaks Spanish, therefore to know this language can open opportunities with Brazilian Scientists, besides Portugal and some African countries has it as official language. Table \ref{tab:port}, reading 2,963 (20.42\%), writing  2,352 (16.22\%), speaking 2,311 (15.93\%).

\begin{table}[hbpt]
\label{tab:port}
\caption{Portuguese Level}
\centering
\begin{tabular}{|c|c|c|c|c|}
\hline
\textbf{Portuguese Level} & \textbf{Reading} & \textbf{Speaking} & \textbf{Writing} & \textbf{Total\_fil} \\ \hline
Basic                     & 1300             & 1911              & 1952             & \textbf{5163}       \\ \hline
Intermediate              & 1512             & 1318              & 1330             & \textbf{4160}       \\ \hline
Advanced                  & 1150             & 772               & 757              & \textbf{2679}       \\ \hline
Superior Advanced         & 301              & 262               & 224              & \textbf{787}        \\ \hline
\textbf{Total\_col}       & \textbf{4263}    & \textbf{4263}     & \textbf{4263}    &                     \\ \hline
\end{tabular}
\end{table}

Checking other languages, i.e. Frech, Italian. The numbers are(reading/speaking/writing): 1,205 (8.31\%)/944(6.51\%)/919(6.34\%) and 601(4.14\%)/454(3.13\%)/443(3.05\%). See tables \ref{tab:fre}, \ref{tab:ita}.

\begin{table}[hbpt]
\label{tab:fre}
\caption{French Level}
\centering
\begin{tabular}{|c|c|c|c|c|}
\hline
\textbf{French Level} & \textbf{Reading} & \textbf{Speaking} & \textbf{Writing} & \textbf{Total\_fil} \\ \hline
Basic                 & 672              & 933               & 958              & \textbf{2563}       \\ \hline
Intermediate          & 587              & 453               & 484              & \textbf{1524}       \\ \hline
Advanced              & 420              & 309               & 276              & \textbf{1005}       \\ \hline
Superior Advanced     & 198              & 182               & 159              & \textbf{539}        \\ \hline
\textbf{Total\_col}   & \textbf{1877}    & \textbf{1877}     & \textbf{1877}    &                     \\ \hline
\end{tabular}
\end{table}

\begin{table}[hbpt]
\label{tab:ita}
\caption{Italian Language}
\centering
\begin{tabular}{|c|c|c|c|c|}
\hline
\textbf{Italian Level} & \textbf{Reading} & \textbf{Speaking} & \textbf{Writing} & \textbf{Total\_fil} \\ \hline
Basic                  & 449              & 596               & 607              & \textbf{1652}       \\ \hline
Intermediate           & 373              & 287               & 290              & \textbf{950}        \\ \hline
Advanced               & 165              & 116               & 105              & \textbf{386}        \\ \hline
Superior Advanced      & 63               & 51                & 48               & \textbf{162}        \\ \hline
\textbf{Total\_col}    & \textbf{1050}    & \textbf{1050}     & \textbf{1050}    &                     \\ \hline
\end{tabular}
\end{table}

\subsubsection{Peruvian Languages}
\label{subsubsec:per}

After searching other kind of languages in Peru, two main ancient languages are still present. Quechua, Aymara both are speaking in Peru today. Quechua has more than 600 people who can read, speak, write with a level higher than intermediate(Table \ref{tab:que}), this number is distributed in 4.45\%/ 4.54\%/ 3.62\%. And Aymara(see Tab. \ref{tab:aya}) has less speakers, a number around 60 and the distribution is 68(0.47\%), 76(0.52\%), 59(0.41\%). By consequence, this language can be extint soon if it is not preserved and promoted from national government.

\begin{table}[hbpt]
\label{tab:que}
\caption{Quechua}
\centering
\begin{tabular}{|c|c|c|c|c|}
\hline
\textbf{Quechua Level} & \textbf{Reading} & \textbf{Speaking} & \textbf{Writing} & \textbf{Total\_fil} \\ \hline
Basic                  & 369              & 357               & 490              & \textbf{1216}       \\ \hline
Intermediate           & 323              & 293               & 300              & \textbf{916}        \\ \hline
Advanced               & 261              & 284               & 169              & \textbf{714}        \\ \hline
Superior Advanced      & 62               & 81                & 56               & \textbf{199}        \\ \hline
\textbf{Total\_col}    & \textbf{1015}    & \textbf{1015}     & \textbf{1015}    &                     \\ \hline
\end{tabular}
\end{table}

\begin{table}[hbpt]
\label{tab:aya}
\caption{Aymara Language}
\centering
\begin{tabular}{|c|c|c|c|c|}
\hline
\textbf{Aymara Level} & \textbf{Reading} & \textbf{Speaking} & \textbf{Writing} & \textbf{Total\_fil} \\ \hline
Basic                 & 31               & 23                & 40               & \textbf{94}         \\ \hline
Intermediate          & 29               & 26                & 30               & \textbf{85}         \\ \hline
Advanced              & 31               & 37                & 20               & \textbf{88}         \\ \hline
Superior Advanced     & 8                & 13                & 9                & \textbf{30}         \\ \hline
\textbf{Total\_col}   & \textbf{99}      & \textbf{99}       & \textbf{99}      &                     \\ \hline
\end{tabular}
\end{table}

Finally, Peruvian community has the potential of growing and leverage his research level and possibly, create technology. More than half holds a Master Degree, and a third one Phd. Most of them are working in National Universities and Institutes, then have the place to develop research groups, collaboration. Besides, has the access to Science through English, and Portuguese, French, Italian to open international projects, conferences and more. By other hand, Peruvian ancient languages requires preservation and can be the start of many studies, all this effort to integrate national community and foster studies from many fields, i.e. Social Sciences, Linguistics, Engineering and more.

\section{Conclusions}

The Text Mining tasks involves many steps, from capture, cleaning, processing to visualization. It is important to notice that real data is not clean, follows a specific format or even uses the same language. 
Filtering data and organizing it, can help understand how is Peruvian situation and how professionals are ready to grow up and evolve in Industry/Academy. Most of them are working in National Academic Institutions, then it must be a good place to learn. collaborate and promote Science.
At the end, languages are a key to connect with other countries, continents and create, organize international projects where researchers can collaborate/support each other and build a community to promote Science, Technology and progress to their countries.

\section{Considerations}

This lines are suggestions to improve your work:
\begin{itemize}
    \item First, if you want to research and answer a question, maybe to solve a problem, you need data to answer and support the analysis.
    \item If data is not available, check alternative sources, i.e. you want to analyze government policies but there is no data, maybe it is possible to get data from other similar countries with policy of open data and it is possible to do a extrapolation or you can find social data to analyze the impact over the population. This step is key to develop next steps.
    \item Real data is not clean, clear for exploration then you are going to invest time, reading, testing ideas, cleaning until this data is ready, besides you can lose data, i.e. 20\% of the dataset. This evaluation must be think and considered for next steps.
    \item Finally, maybe there is no specific data about your question, then you need to create artificial variables to understand the data, this involves much creativity and imagination. Remember, it is possible to cross dataset to get more meaningful data.
\end{itemize}

\section{Future Work}

The authors explore Peruvian situation considering only 25000 curriculum vitae, the next step is to replicate the analysis over the entire existent registers. And, create a tool to support and foster collaboration projects between Peruvians and foreigners to catapult Science in Latin America.

\section*{Acknowledgments}
This final sections is to thank Concytec for the role of promoting Science, Technology in Peru. A country supported by Science and Technology can be a sustainable nation in the short future and influence in South America region. Finally, the authors want to mention Research4Tech, an Artificial Intelligence community of Latin American Researchers for promoting Science and collaboration in Latin American countries, his roles as integrator between Professional, Researchers, Technology communities is key to develop the Latin American region as a strong body.

\bibliographystyle{IEEEtran}
\bibliography{biblio}

\end{document}